\documentclass[journal]{IEEEtran}
\usepackage{amsmath,mathtools,amsfonts,amssymb,amsthm}
\usepackage{float}
\usepackage{algpseudocode}
\usepackage[ruled,vlined]{algorithm2e}
\usepackage{array}
\usepackage[caption=false,font=normalsize,labelfont=sf,textfont=sf]{subfig}
\usepackage{textcomp}
\usepackage{stfloats}
\usepackage{url}

\usepackage{url}
\usepackage{xcolor}
\usepackage{verbatim}
\usepackage{graphicx}
\usepackage{soul}
\def\BibTeX{{\rm B\kern-.05em{\sc i\kern-.025em b}\kern-.08em
    T\kern-.1667em\lower.7ex\hbox{E}\kern-.125emX}}
\usepackage{balance}
\usepackage{multirow}
\usepackage{booktabs} 

\usepackage{caption}                         
 \usepackage{multirow}
 \usepackage{threeparttable}
\usepackage{mathtools}  
\PassOptionsToPackage{linesnumbered,ruled,vlined}{algorithm2e}
\usepackage{tablefootnote}
\usepackage[numbers,sort&compress]{natbib}

\newtheoremstyle{sltheorem}
{}                
{}                
{}        
{10pt}                
{\bfseries}       
{:}               
{ }               
{}                
\theoremstyle{sltheorem}

\DeclareMathOperator*{\argmax}{arg\,max}

\begin{document}

\title{Modeling Coincident Peak Pricing in Electricity Markets: Challenges and Peak Shaving Effectiveness}

\author{Qian Zhang, Sadie Zhao, Lucy Diao, Conleigh Byers, Yiling Chen, Derya Cansever, Le Xie
\thanks{Qian Zhang, Sadie Zhao, Yiling Chen, Derya Cansever, and Le Xie are with John A. Paulson School of Engineering and Applied Sciences, Harvard University, USA. Lucy Diao is with the College, the University of Chicago, USA. Conleigh Byers is with Harvard Kennedy School of Government, Harvard University, USA. (correspondence e-mail: xie@seas.harvard.edu).}
}



\maketitle

\begin{abstract}
Coincident Peak (CP) pricing is widely used in U.S. electricity markets to allocate capacity and transmission costs. This paper develops a behavioral game-theoretic framework for CP-driven load shifting that couples a nonlinear cost-allocation model with day-ahead (one-shot) and real-time (sequential-learning) decision processes. We examine two update rules, namely best-response dynamics (BRD) and fictitious-play dynamics (FPD), across continuous and finite action spaces to quantify how flexibility, action resolution, and participation influence peak outcomes. Using ERCOT peak-day data, we find that FPD reliably reduces system peaks, whereas BRD is more variable and can \emph{increase} peaks under tight-capacity conditions. Finer action resolution improves peak shaving, while the number of participants is largely neutral when aggregate flexibility is fixed. 
Meanwhile, information-provider signals can induce herding, whereas response-aware or diverse signals improve peak shaving. These results highlight both the potential and limits of CP pricing: smoothing information and enabling granular control are as important as the amount of available flexibility. The framework offers practical guidance for system operators and consumers: \emph{For ISOs}, broadcasting smoothed CP signals and setting minimum controllable-capacity thresholds enhance coordination. \emph{For consumers}, greater flexibility and finer control resolution improve both cost savings and peak-shaving performance.
\end{abstract}


\section{Introduction}

As new large-scale electricity demands, such as data centers, continue to grow, the U.S. power grid requires substantial generation and transmission expansion to meet future needs. Unlike the short-term energy market, long-term planning and investment cost allocation involve greater uncertainty and present additional challenges in fairly distributing costs. A key difficulty lies in defining the notion of ``benefit'' while ensuring that the allocation method remains computationally tractable, based on measurable quantities, and does not distort market signals. 



In this paper, we focus on cost allocation based on \emph{Peak Load Contribution} (PLC), also referred to as \emph{Coincident Peak} (CP), which measures a customer’s or Load Serving Entity’s (LSE) electricity demand during periods of highest system-wide demand. Owing to its transparency, simplicity, and ease of implementation, CP has been widely adopted by Independent System Operators (ISOs) in the U.S.~\cite{baldick2018incentive}, and has become a primary metric for allocating capacity and transmission costs in many markets. For example, ISO New England allocates capacity charges via 1~CP from June to May based on metered or estimated demand data reflecting hourly integrated electric consumption~\cite{ISO-NE2023}. For systems without a capacity market, CP is primarily used to allocate transmission investment costs. For instance, the Electric Reliability Council of Texas (ERCOT) uses 4~CP measurements taken from the four highest demand hours during June to September to determine each customer's share of transmission charges~\cite{hogan2017priorities,du2019demand}.

Although the CP method is widely accepted, it still poses several potential market distortions. One issue lies in determining the appropriate measurement frequency. For example, in CAISO, the 12CP approach is considered more effective in reflecting both how the transmission system is planned and how customers utilize and benefit from transmission service~\cite{CAISO2018}. Another challenge, especially with increasing demand flexibility, concerns the peak-shaving effect of CP; first, whether CP is effective at co-incident demand reduction, and second, whether such reduction is desirable from a market efficiency perspective. When used to recover sunk costs, CP may introduce market distortions and cost allocation biases, particularly in energy-only markets such as ERCOT~\cite{hogan2017priorities}. Allocating sunk costs to variable energy costs causes an asymmetry in the cost a consumer faces (energy price + capacity charge) and the price a supplier is paid (energy price) during CP events. This distorts efficiency in the energy market, reducing the ability of producers to recover costs in the energy market due to lower prices and reducing welfare for consumers who would prefer to consume at the prevailing energy price. This may lead to under-investment in generation. CP may also shift sunk costs to inflexible consumers, which could be socially undesirable, particularly if these are small-scale commercial and residential consumers. For new transmission investments that could be avoided by reducing peak demand growth, CP may provide a coordinating signal in the absence of ideal time-varying energy prices \cite{baldick2018incentive}. However, new transmission investments are often substitutes for new generation, which could improve reliability at a lower cost if given the appropriate energy price signal to enter the market. This paper addresses the first question of whether CP reliably incentivizes peak shaving, leaving the second question of the welfare impacts of peak shaving to future work.


Before evaluating the market-level impact of the CP mechanism, it is important to first understand the key factors and challenges in modeling user behavior under CP. 
First, the peak interval is typically defined based on system-level demand, which poses significant challenges for forecasting due to the presence of massive, heterogeneous consumers. Second, the ISO and third-party companies also publish their demand forecasts, which can be considered as shared information available to market participants, adding complexity to the modeling of their behavior \cite{ERCOT_LoadForecast_CurrentDay}. 
Finally, CP is typically regarded as a demand response program to reduce the peak demand, and it often overlaps with other demand response programs \cite{Ogelman2023}.

Over the past decade, many researchers have sought to understand forecasting and user behavior under CP. 
The challenge of CP prediction was examined in \cite{jiang2014predicting,jiang2016analyzing}, which found that even state-of-the-art peak prediction algorithms require consumers to curtail load ten or more times to avoid certain peaks. 
A more advanced scenario generation and Monte Carlo estimation method was proposed in \cite{carmona2024coincident} for CP prediction. 
The game-theoretic framework was proposed in \cite{chen2025gaming} to model the interactions among consumers and the related Nash equilibrium properties under the fixed CP price assumption. Relatedly, evolutionary game-theoretic perspectives have been used to study strategy evolution and coordination in sustainable energy markets and decentralized power systems \cite{cheng2025jclepro_egt,cheng2025rser_gt}. On the other hand, some studies focus on analyzing the peak-shaving capability of different types of demand, which is crucial for system operators when planning future resources and transmission lines \cite{zarnikau2013response}. 
For example, \cite{brannlund2021peak} reveals the limited peak-shifting capability in Sweden, \cite{kim2025estimation} evaluates the peak demand reduction potential using smart thermostats in ERCOT, and \cite{wu2022design} develops a battery energy management system and dispatch algorithm to reduce the system peak.

\emph{Contributions.} This paper makes four contributions. First, we formalize a nonlinear CP cost-allocation model and a tractable behavioral load model for responsive demand with capacity and energy-balance constraints under both continuous and finite action sets. Second, we link day-ahead one-shot decisions with real-time sequential learning and analyze two canonical response dynamics, highlighting key theoretical and computational challenges. 
Third, we incorporate the role of information providers and show how it affects peak-shaving performance, including the system-level value of heterogeneous signals.
Finally, using ERCOT peak-day data, we quantify how flexibility, action resolution, participation, and information diversity shape outcomes. These findings yield actionable guidance for both system operators and participants: for ISOs, broadcasting smoothed CP-likelihood signals and setting minimum flexibility thresholds enhance peak-shaving effectiveness, while for responsive consumers, maintaining adequate flexible capacity, adopting finer control resolution, and leveraging diversified forecasting signals improve peak avoidance.

The remainder of this paper is organized as follows. Section~\ref{sec:cost} introduces the CP cost-allocation model and its linear approximation. Section~\ref{sec:user} formalizes consumer constraints and describes decision-making in the day-ahead and real-time markets. Section~\ref{sec:game} develops the game-theoretic formulations for both settings and highlights key analytical challenges. Section~\ref{sec:nash} evaluates peak-shaving effectiveness using counterexamples and practical insights. Section~\ref{sec:case} presents case studies based on ERCOT peak-day data under both continuous and finite action spaces. The code of this paper is
available at \url{https://github.com/qian-harvard/Coincident-Peak-Game}

\section{Cost Allocation Model} \label{sec:cost} 
\subsection{Preliminary}
We consider a time horizon consisting of $T$ discrete intervals, denoted by the set $\mathcal{T} = \{1, 2, \dots, T\}$. The duration of these intervals (e.g., 5 minutes or hourly) is determined by the specific market rules of the system operator.

Large consumers who pay CP are divided into two groups: responsive consumers who can shift demand from one interval to another, and non-responsive consumers whose demand is assumed to be inelastic. Let $x_i(t)$ be the demand of a responsive consumer $i$ at time $t$,  for $t \in \mathcal{T}$ and $i \in \{1, \dots, N\}$.

Since the focus of this work is on modeling the game and behavior of the responsive load, the non-responsive consumers are aggregated. The aggregate demand of the non-responsive consumers at time $t$ is given by $B(t)$. Consequently, the total system demand at time $t$ can be expressed as:
\begin{equation}
S(t) = B(t) + \sum_{i=1}^N x_i(t)
\end{equation}

The primary objective of the cost allocation model is to distribute the total system cost \( C \) among electricity consumers. This cost typically includes expenditures related to future transmission system maintenance and infrastructure investment. In systems with capacity markets, it also includes the capacity payments intended to ensure long-term resource adequacy. For simplification, this study focuses exclusively on the transmission component of the cost, as is the case in markets like ERCOT.

\subsection{Nonlinear Cost Allocation Model}

Let $t^* = \arg\max_{\tau \in \mathcal{T}} S(\tau)$ 
denote the system coincident peak interval. The transmission charge allocated to a responsive consumer \( i \) is given by  
\begin{equation} \label{eq:costallocate_single}
c_i = C \cdot \frac{x_i(t^*)}{S(t^*)},
\end{equation}  
where \( C \) is the total transmission cost and \( S(t) \) is the aggregate system demand at time \( t \).  

\emph{Remark:} If multiple time intervals share the same maximum demand, the transmission charge is proportionally allocated to customers based on their \emph{total} consumption during these peak intervals:  
\begin{equation} \label{eq:costallocate_multi}
c_i = C \cdot \frac{\sum_{t \in \mathcal{T}} \mathbb{I}_{t \in \arg\max_{\tau \in \mathcal{T}} S(\tau)} \, x_i(t)}
{\sum_{t \in \mathcal{T}} \mathbb{I}_{t \in \arg\max_{\tau \in \mathcal{T}} S(\tau)} \, S(t)},
\end{equation}  
where \( \mathbb{I}_{t \in A} \) is an indicator function that equals 1 if \( t \) belongs to set \( A \), and 0 otherwise.  

This cost allocation model is highly nonlinear. The individual charge \( c_i \) depends on the consumer's own demand profile \( x_i(t) \), which appears in both the numerator and denominator through the total system load \( S(t) = B(t) + \sum_{k=1}^N x_k(t) \). This mutual dependence between individual and aggregate behavior introduces nonlinearity and plays a central role in shaping their strategic responses under the CP pricing mechanism.

\subsection{Linearization and Limitation}

The nonlinear structure of the cost allocation model introduces challenges for optimization and strategic behavior analysis. To simplify analysis, linearized approximations isolate the influence of each user's contribution at the system peak. Let \( x_{i,0}(t^*) \) be a reference demand for user \( i \), \( S_0 = \sum_{j=1}^N x_{j,0}(t^*) \), and \( D_0 = B(t^*) + S_0 \). One simplified formulation based on Taylor expansion expresses the charge to user \( i \) at the peak time $t^*$ as:

\begin{equation}
\hat{c}_i = k_i \cdot x_i(t^*) + d_i
\end{equation}
where
\begin{equation}
k_i = \frac{C}{D_0^2} \left( D_0 - x_{i,0}(t^*) \right)
\end{equation}
and
\begin{equation}
d_i = \frac{C}{D_0^2} [ - x_{i,0}(t^*) \sum_{k \neq i} x_k(t^*) + x_{i,0}(t^*) S_0 ]
\end{equation}

When the responsive portion of demand is relatively small compared to the baseline load, the interaction between users becomes negligible, i.e. $B(t) \gg \sum_{i=1}^N x_i(t)$. In this case, $d_i \approx \frac{C}{B(t^*)^2}\,x_{i,0}(t^*)^2$ and the model can be further simplified to a fixed-price approximation:
\begin{equation}
c_i \approx x_i(t^*) \frac{C}{B(t^*)} \
\end{equation}
In this expression, the term \( \frac{C}{B(t^*)} \) can be interpreted as an effective fixed capacity charge price during the identified peak period \cite{chen2025gaming}. While the linearized model improves tractability, it overlooks key nonlinear interactions. It assumes individual demand has minimal impact on the system peak, which may not hold in systems with high flexibility. As a result, it may misrepresent incentives and overlook strategic behavior.

\section{Load Behavior Assumption in Electricity Markets} \label{sec:user}

Consumers exhibit diverse demand response behaviors, which are challenging to capture comprehensively in a single model. This section focuses on the common constraints that shape user behavior in response to the coincident peak pricing mechanism.

\subsection{Load Behavioral Constraints}

\subsubsection{Capacity Constraints} \quad

Each responsive consumer is subject to lower and upper bounds on their demand at any given time, reflecting their individual flexibility levels:
\begin{equation}
\underline{X}_i \le x_i(t) \le \overline{X}_i
\end{equation}

\subsubsection{Energy Balance Window} \quad

The total energy consumed by load \( i \) over a designated time window \( \mathcal{W} := \{t_1, t_2, \ldots, t_W\} \) is assumed to be fixed at a constant value \( E_i \). This captures the energy balance constraint:
\begin{equation} \label{ebc}
\sum_{t \in \mathcal{W}} x_i(t) = E_i
\end{equation}
The size of the energy balance window depends on the temporal shifting capability of different types of flexible demand. For example, some heat pumps may offer intra-hour flexibility, while electric vehicle (EV) charging may allow shifting over several hours.

\subsubsection{Reformulation} \quad

To better understand the shifting behavior of consumers, we introduce a baseline consumption profile in a time window period $\mathcal{W}$. Assume the reference electricity consumption profile of consumer $i$ is: $\{ b_i(t_1), b_i(t_2), ..., b_i(t_W) \} $. It is clear that it meets the energy balance constraint:
\begin{equation}
\sum_{t \in \mathcal{W}}  b_i(t) = E_i
\end{equation}

Then, we introduce the action $\{a_i(t_1), ..., a_i(t_W) \}$ to represent users' shifting behavior,  we have
\begin{equation}
x_i(t) = b_i(t) + a_i(t)
\end{equation}
where $\sum_{t \in \mathcal{W}}  a_i(t) = 0$ because of energy balance constraint. 

\emph{Remark:} Other constraints, such as shifting penalties and ramping limits, are omitted in this study for simplicity. This simplification is justified by two factors: (1) the CP charge is typically much higher than conventional energy prices, providing strong incentives for demand response and allowing the omission of shifting penalties; and (2) many emerging flexible resources, such as bitcoin mining facilities and battery storage systems, possess substantial ramping capabilities.

\subsection{Electricity Markets}

Although peak demand is strongly correlated with temperature, forecasting the exact timing of the system-wide coincident peak has become increasingly difficult in ERCOT since 2024. This is due to the growing prevalence of battery storage and the increasing number of consumers actively responding to CP pricing signals. As more users engage in strategic demand shifting, the CP peak becomes an endogenous outcome of aggregate user behavior rather than a deterministic forecast.

Before introducing the game theoretical model, we briefly discuss the different decision-making processes in the day-ahead market and real-time market. We assume that the peak interval is known only within a single energy-balance window, i.e.,$t^* = \arg\max_{\tau \in \mathcal{T}} S(\tau) \in \mathcal{W}$, which is taken to represent one day in this study. This assumption is typically valid in practice, as monthly peak day forecasting tends to be reasonably accurate. Users can thus plan their demand-shifting strategies with the expectation that the peak will occur within a specific, limited set of hours.

\subsubsection{Decision Making in the Day-Ahead Market} \quad

In the day-ahead market, users may adjust their demand profiles over the 24-hour horizon in anticipation of potential CP intervals. In practice, multi-round bidding is not a standard procedure due to concerns such as market collusion and timing constraints \cite{contreras2002auction}. Consequently, only a single-round bid is permitted in the day-ahead market, resulting in a one-shot game formulation that poses analytical challenges, as discussed in the following section.

Nevertheless, the day-ahead submission establishes the baseline schedule for subsequent real-time adjustments. Within our framework, the day-ahead bid can serve as the initial condition for the rolling real-time decision process, constraining each agent's feasible actions through energy balance and capacity limits. Since the actual CP is determined by real-time system demand (typically measured at 5- or 15-minute intervals), CP-responsive behavior in the day-ahead stage exerts only an indirect influence on the realized CP. Even so, the day-ahead peak hour forecasting provides informative signals for identifying candidate real-time peak periods, thereby narrowing the search space for response and guiding subsequent best-response or fictitious-play updates within the rolling decision window.

\subsubsection{Decision Making in the Real-Time Market} \quad

Since the CP interval is ultimately determined by the system’s real-time electricity consumption at the end of the month, user behavior in practice tends to be dynamic and adaptive. Participants continually update their strategies in response to historical consumption data and observed actions of other users, typically in a sequential rather than iterative manner. 


Modeling this dynamic interaction in real time presents several challenges: (1) stochastic and heterogeneous user behaviors; (2) a high-dimensional action space induced by temporal coupling under energy-balance constraints; and (3) strategic interdependence stemming from reliance on similar forecasting information sources, which often precludes closed-form equilibrium solutions.

\emph{Summary and Key Assumptions:} CP charges are ultimately based on \emph{metered real-time physical load} at the coincident peak; therefore, day-ahead financial positions affect CP exposure only if they change real-time dispatch or consumption. In addition, system peaks often occur during high real-time price periods, so real-time energy-cost minimization may partially align with CP shaving, although this alignment is not guaranteed under forecast error, binding capacity and energy-balance constraints, and the non-additive CP allocation. Motivated by these observations, we model the day-ahead stage as a one-shot baseline schedule and focus on rolling-window real-time updates as the primary channel through which participants adapt consumption to manage CP risk.

The next section formulates the game-theoretic representations for both the day-ahead and real-time markets. For the day-ahead formulation, we address the primary analytical difficulties, while for the real-time setting, we retain a minimal yet sufficient structure to capture the adaptive learning process exhibited by consumers.

\section{Game Theoretical Model}\label{sec:game}
\subsection{One-shot Strategic Interaction: The Load Consumption Game in Day-Ahead Market}
To delve into consumers' strategic interactions under the presence of the CP mechanism in the day-ahead market, we formulate it as a normal-form \textit{Load Consumption Game} $(N, (\mathcal{A}_i)_{i\in N}, (u_i)_{i\in N})$ where 
\begin{itemize}
    \item $N$ is the set of responsive consumers,
    \item For each $i\in N$, the action space follows the load behavioral constraints, i.e.
    \begin{align*}
        \mathcal{A}_i=\bigg\{&\mathbf{a}_i = (a_i(t))_{t\in \mathcal{W}}\mid 
        \sum_{t\in \mathcal{W}}a_i(t)=0, \\
        &\forall t\in \mathcal{W}, 
        \underline{X}_i \leq  b_i(t) + a_i(t)  \leq \overline{X}_i\bigg\}
    \end{align*}
    which can be either a continuous or a finite action space.
    \item For each $i\in N$, we model the agent's objective as minimizing a \emph{total cost} consisting of the CP allocation charge and energy payments. Equivalently, define
    \begin{align*}
        J_i(\mathbf{a}_i, \mathbf{a}_{-i})
        = &C\frac{\sum_{t \in \mathcal{W}} \mathbb{I}_{t \in \arg\max_{\tau \in \mathcal{W}} S(\tau)} \, b_i(t)+a_i(t)}{\sum_{t \in \mathcal{W}} \mathbb{I}_{t \in \arg\max_{\tau \in \mathcal{W}} S(\tau)} \, S(t)}\\
        &+ \sum_{t\in \mathcal{W}} \pi_t(b_i(t)+a_i(t))
    \end{align*}
    where $\pi_t$ is the energy price at time $t$. And we can set $u_i = -J_i$ for the game-theoretic notation.
\end{itemize}

\emph{Price-taking Assumption}: In our analysis, $\pi_t$ is treated as exogenous (price-taking loads), so strategic coupling arises primarily through the CP term via the endogenous peak $t^*$.

To study the strategic interactions among responsive consumers, a natural starting point is the concept of a \textit{Nash equilibrium}. If we assume that each agent has a \emph{finite} action space, the \textit{Load Consumption Game} is guaranteed to admit at least one \textit{mixed-strategy Nash equilibrium}. Nevertheless, analyzing Nash equilibria in the day-ahead market directly is not an ideal approach in this scenario for two main reasons:
\begin{itemize}
    \item \textbf{From a behavioral and informational perspective}, the Nash equilibrium presumes that all consumers best respond to each other simultaneously, which in turn requires each consumer to possess complete knowledge of other consumers’ action spaces and utility functions. In practice, however, consumers in electricity markets operate under \emph{asymmetric} and \emph{limited information}.
    \item \textbf{From a computational perspective}, computing a mixed Nash equilibrium in a general finite game is \emph{PPAD-hard}, implying that no efficient (polynomial-time) algorithm is known for finding one. Consequently, equilibrium computation can become \emph{intractable} in large-scale systems involving many consumers.
\end{itemize}

\subsection{Dynamic Strategic Interaction: Learning in Real-Time Markets}
While directly computing a Nash equilibrium may be impractical, consumers in electricity markets naturally have opportunities to \emph{update} their energy consumption sequentially over the course of real-time market operations. This sequential interaction provides a natural mechanism for consumers to \emph{adapt their strategies in response to others’ observed behaviors}, leading to an emergent adjustment process rather than an instantaneous equilibrium computation.

In this subsection, we formalize this dynamic interaction among consumers through two well-established game-theoretic learning frameworks: \textbf{best-response dynamics} and \textbf{fictitious play dynamics}.
In \textit{best-response dynamics}, at each iteration (or market round), consumers myopically choose their strategy that best responds to the current actions of others, assuming those actions remain fixed.
In contrast, under \textit{fictitious play dynamics}, each consumer forms a belief about others’ strategies based on the empirical frequency of their past actions, and best responds to this belief.
Both dynamics capture how consumers may gradually adjust their behavior over time, providing a more realistic model of strategic adaptation in repeated or sequential market environments.

Consider another time window $\mathcal{H} := \{h_0, h_1, \ldots, h_H\} \subseteq \mathcal{T}$, which may have a different duration than $\mathcal{W}$.
At each $h \in \mathcal{H}$, a day-ahead or real-time market takes place, and each responsive consumer $i \in N$ can submit their bid $\mathbf{a}_i^h \in \mathcal{A}_i$.
However, consumers cannot revise their consumption for time periods that have already been realized. Formally, for any $t \in \mathcal{T}$ and all $h \ge t$, we have $\mathbf{a}_i^h(t)=a_i^{\dagger}(t)$, where $a_i^{\dagger}(t)$ denotes the final bid submitted for period $t$ at the last opportunity before its realization (i.e., the largest $h < t$).
Therefore, for every $h\in \mathcal{H}\setminus\{h_0\}$, we further define the \textit{feasible action space with freezing constraint} as: 
\begin{align*}
    \mathcal{A}_i^h =\{\mathbf{a}_i^h\in \mathcal{A}_i\mid \mathbf{a}_i^h(t)=a_i^{\dagger}(t)\;\forall t\leq h\}
\end{align*}

Now, we can formalize the \emph{best-response dynamics} and \emph{fictitious play dynamics}:
\subsubsection{Best-response Dynamics} 
\begin{itemize}
    \item When $h=h_0$, each consumer $i\in N$ pick an arbitrary action $\mathbf{a}_i^0 \in \mathcal{A}_i$ or a user-defined baseline action.
    \item For any subsequent market time $h=t_\tau$ with $\tau\in \{1, \cdots, H\}$, each consumer $i\in N$ picks action
    \begin{equation} \label{brd}
    \begin{aligned}
        \mathbf{a}_i^\tau\in \argmax_{\mathbf{a}_i\in \mathcal{A}_i^\tau} u_i(\mathbf{a}_i, \mathbf{a}_{-i}^{\tau-1})
    \end{aligned}        
    \end{equation}

\end{itemize}
\subsubsection{Fictitious Play Dynamics}
\begin{itemize}
    \item When $h=h_0$, each consumer $i\in N$ pick an arbitrary action $\mathbf{a}_i^0 \in \mathcal{A}_i$ or a user-defined baseline action.
    \item For any subsequent market time $h = h_\tau$ with $\tau \in \{1, \ldots, H\}$, each consumer $i \in N$ first computes the \textbf{empirical frequency} of every other consumer $j \in N \setminus \{i\}$’s past actions up to $h_{\tau-1}$:
\begin{align*}
    p_j^{\tau-1}(\mathbf{a}_j)
    = \frac{1}{\tau} \sum_{k=0}^{\tau-1} 
    \mathbb{I}\{\mathbf{a}_j^k = \mathbf{a}_j\},
    \quad \forall\, \mathbf{a}_j \in \mathcal{A}_j.
\end{align*}
where the denominator $\tau$ represents the total number of observed periods up to $h_{\tau-1}$. These empirical frequencies are combined into a joint belief over opponents’ actions:
\begin{align*}
    p_{-i}^{\tau-1} = \bigotimes_{j \neq i} p_j^{\tau-1}.
\end{align*}

 Finally, consumer $i$ best responds to this joint belief by choosing:
 \begin{equation} \label{fpd}
\begin{aligned}
    \mathbf{a}_i^\tau
    \in \arg\max_{\mathbf{a}_i \in \mathcal{A}_i}
    \mathbb{E}_{\mathbf{a}_{-i} \sim p_{-i}^{\tau-1}}
    \big[\,u_i(\mathbf{a}_i, \mathbf{a}_{-i})\,\big].
\end{aligned}   
\end{equation}

\end{itemize}

While obtaining theoretical results from the above models is challenging, valuable insights can still be gained by focusing on specific system properties, particularly the effectiveness of CP pricing in promoting \emph{peak shaving}. Before conducting some case studies in Section \ref{sec:case}, several simplified yet counterintuitive examples in Section \ref{sec:nash} provide meaningful intuition for evaluating CP-based demand response mechanisms.

\section{Peak Shaving Effectiveness} \label{sec:nash}

From a policy standpoint, as part of demand response programs administered by ISOs such as ERCOT, it is crucial to assess how effectively CP pricing mechanisms achieve their intended objective of reducing system peak demand~\cite{du2019demand}.  

\subsection{Counterexample and Key Factors in Peak Shaving}

The CP mechanism has demonstrated substantial peak reduction in ERCOT~\cite{Ogelman2023}, though most findings are based on posterior statistical inference. As the share of flexible demand remains relatively small in most power systems, such evidence primarily provides insights at the macro level.

To evaluate the effectiveness of peak shaving, the outcomes must be compared against the initial system state prior to the CP game. This baseline includes not only the non-responsive demand profile but also the initial demand schedules of flexible consumers. We revisit the peak shaving effectiveness of CP pricing from a game-theoretic perspective with some counterintuitive examples, where users may fail to coordinate in a way that reduces the system peak. 

To illustrate the potential failure of peak shaving in a one-shot game, assume that energy prices are identical across intervals and that players apply a \emph{pure strategy} within a \emph{finite} action space to avoid the coincident peak. Consider a two-player, two-period game with a coarse action space:
\[
\mathcal{A}_i = \{ (0, E_i), (E_i, 0) \},
\]
which restricts each player to allocate their entire demand to a single period, representing an extreme form of flexibility.  

Let \( E_1 = E_2 = E \), \( B(t_1) = B(t_2) \), and assume no additional capacity constraints. If Player~1 chooses \( (E,0) \) and Player~2 chooses \( (0,E) \), the system load becomes
\[
S(t_1) = E + B(t_1), \quad S(t_2) = E + B(t_2),
\]
resulting in a flat profile where no player has an incentive to deviate.  

However, under a limited action space or starting from improper initial states, it might fail to achieve an equilibrium that flattens the curve under the best response assumption. For example, if both players initially choose \( (E,0) \), then
\[
S(t_1) = 2E + B(t_1), \quad S(t_2) = B(t_2),
\]
making \( t_1 \) the peak. Each player, therefore, has an incentive to shift to the opposite interval in a one-shot game, preventing convergence to a stable equilibrium.

The above extreme case demonstrates that the effectiveness of peak shaving depends on multiple factors and assumptions. Even under the same baseline, variations in user behavior can yield different outcomes. In particular, the set of available load-shifting actions and the choice of strategies play critical roles in determining the extent of peak reduction. Nevertheless, the CP mechanism is often effective under more realistic conditions, where action spaces are richer and players adopt more adaptive strategies. Peak shaving is particularly evident when:

\begin{itemize}
    \item \textbf{The action space is fine-grained}: Smaller step sizes (i.e., more continuous control over \( x_i(t) \)) allow users to smooth their load profiles and share the peak more evenly.
    \item \textbf{The energy balance window is large}: A wider window \( \mathcal{W} \) offers more temporal flexibility, allowing peak avoidance without creating new peaks elsewhere.
    \item \textbf{The number of players is large}: As the number of responsive users increases, the impact of any single user on the peak diminishes, reducing incentives for aggressive peak chasing and leading to more stable aggregate behavior.
    \item \textbf{Players are heterogeneous and adaptive}: Diverse and smarter strategies reduce synchronized responses, making the system more stable and improving peak shaving performance.
    \item \textbf{Large gaps exist among non-responsive loads across time periods}: A natural asymmetry in \( B(t) \) helps stabilize the peak, reducing oscillation incentives.
\end{itemize}

\subsection{Information Providers}

With the increasing penetration of intermittent resources and the heterogeneity of market participants’ behavior, peak-demand forecasting has become extremely challenging. In recent years, a growing number of third-party firms have begun offering CP-forecasting services. These information providers typically deliver daily (and intra-day) alerts identifying candidate peak periods for the following day or upcoming hours \cite{gridstatus_ercot_4cp,enverus_ercot_cp_2025,peakpower_coincident_peak_notifications}.

Such signals can simplify behavior modeling for large flexible loads: rather than specifying response models for all other players, an agent can condition its decision on the provider’s ranked list of candidate peak intervals. However, widespread reliance on common forecasts may induce correlated actions and synchronization effects, potentially complicating the evaluation of peak-shaving effectiveness and, in some cases, diminishing system-level benefits. Assessing these equilibrium implications and designing mechanisms that mitigate herding remains an important direction for future work.

\subsection{Non-Responsive Demand}

Unlike other commodity markets, a substantial portion of electricity demand is inelastic, or non-responsive to price signals. The baseline profile of this non-responsive demand significantly influences the behavior of CP-responsive consumers and, consequently, the overall peak-shaving performance. To illustrate its impact, we present a simple example in the following subsection.

\subsubsection*{Problem Setting}

Consider two strategic consumers, indexed by \( i \in \{1, 2\} \), each deciding how to allocate a fixed energy budget \( E_i \) across two time periods \( t \in \{t_1, t_2\} \). The total system load at time \( t \) is:
\[
S(t) = x_1(t) + x_2(t) + B(t)
\]
and the peak period is defined as $t^* = \arg\max_{t \in \{t_1, t_2\}} S(t)$.

Assuming the energy price is the same for both intervals, then each player aims to minimize total transmission charges. We consider 3 representative cases to show the impact of different non-responsive demand profiles on players' behavior and the Nash equilibrium.

\emph{\textbf{Case 1: Balanced Non-responsive Load}}  

Assume \( B(t_1) = B(t_2) \). In this symmetric setting, consider the strategy where each player allocates demand evenly across the two periods, i.e.,  
\( x_i(t_1) = x_i(t_2) = \tfrac{E_i}{2}.\)  
The resulting system loads are \(S(t_1) = S(t_2)\), where the total system load is identical in both periods, and hence no unique peak occurs. Under this allocation, each player’s CP charge is the same across both periods, and there is no unilateral incentive to deviate by shifting demand. Any deviation (e.g., moving load entirely to one period) would strictly increase that player’s cost by raising their share of the coincident peak. In this case, the equilibrium outcome also corresponds to a perfectly flattened load profile.


\emph{\textbf{Case 2: Highly Imbalanced Non-responsive Load}}

Assume \( |B(t_1) - B(t_2)| \ge E_1 + E_2 \). The peak period is determined entirely by the higher background load, independent of player actions. In this scenario, both players will shift as much load as possible to the non-peak period, constrained by their capacity bounds. Since no player can influence the peak outcome, the strategy of off-peak shifting is dominant and forms a Nash equilibrium.

This case illustrates that the existence of a Nash equilibrium does \emph{not} necessarily imply a flattened demand curve, particularly when the non-responsive load is highly imbalanced. 

\emph{Remark}: Because the majority of non-responsive demand typically comes from residential customers, such scenarios raise an additional policy concern: flexible consumers, such as data centers, can easily avoid CP charges, creating a form of discriminatory cost allocation that shifts a share of costs onto less flexible residential users.

\emph{\textbf{Case 3: Mildly Imbalanced Non-responsive Load}}

Assume \( |B(t_1) - B(t_2)| < E_1 + E_2 \). In this intermediate case, the existence of a Nash Equilibrium becomes complex. The strategic interaction becomes unstable: when one player shifts demand away from the peak, the peak may shift to the other period, prompting the other player to counter-shift. This \emph{might} lead to the non-existence of a Nash equilibrium. 

With the increasing penetration of large flexible loads, future power systems are likely to transition from a Case 2–like scenario toward Case 3. While the problem setting is relatively simple, the three cases illustrate that the players' behavior and Nash equilibrium depend critically on:
\begin{itemize}
    \item The imbalance in non-responsive demand \( B(t) \),
    \item The magnitude of each user’s energy budget \( E_i \),
    \item The flexibility bounds \( [\underline{X}_i, \overline{X}_i] \).
\end{itemize}

\section{Case Studies} \label{sec:case}

This section presents a series of case studies using realistic ERCOT data to illustrate the impact of different response strategy assumptions and parameter settings in the CP mechanism. 

\subsection{System Description} 

\begin{figure}[H] 
\centering
  \includegraphics[scale=0.5]{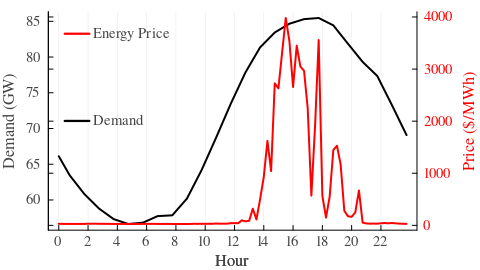}
  \caption{The energy price (red) and demand baseline (black) based on ERCOT 2023 peak day}
  \label{timeslot}
\end{figure}

Figure~\ref{timeslot} presents the energy price and baseline demand data used in this section, derived from the 2023 peak day in the ERCOT real-time energy market~\cite{ercot_rtm}. The total system capacity charge is approximately \$5.72~billion ~\cite{ercot4cp}. Under the CP mechanism, this cost is proportionally allocated to users based on their demand during the system’s four monthly peak intervals. The corresponding capacity charge is approximately \$67{,}000/MWh, an order of magnitude higher than the energy price, shown by the red curve in Figure~\ref{timeslot}.

As reported by ERCOT, the total capacity of \emph{Large Flexible Loads} (LFLs) was approximately 4{,}479~MW in 2023, with most of these loads exhibiting active demand responsiveness~\cite{EIA2024}. Based on this observation, we assume that the aggregate responsive demand is initially flat at 5{,}000~MW prior to any CP-induced adjustments. Each LFL is considered identical, such that
\[
b_i(t) = \frac{5000}{N}, \quad \forall t \in \mathcal{W},
\]
\[
\sum_{i=1}^N b_i(t) = 5000, \quad \forall t \in \mathcal{W},
\]
where \( \mathcal{W} \) denotes the 24-hour energy-balance window. The lower bound of each LFL is set to zero, i.e., \( \underline{X}_i = 0, \ \forall i \).

Accordingly, the aggregate non-responsive demand at time \( t \) is expressed as
\[
B(t) = S(t) - \sum_{i=1}^N b_i(t),
\]
where \( S(t) \) represents the total real-time system demand shown in Figure~\ref{timeslot}.

\emph{Remark:} Our simulation framework centers on the behavior of responsive demand (LFLs) and their contribution to peak shaving under the defined load-behavioral constraints. Although historical demand data have embed partial responses to past CP signals, we assume these effects are minor and do not materially alter the shape of the aggregate non-responsive demand. Accordingly, the observed baseline demand is treated as a representative proxy for the underlying non-responsive component, allowing us to isolate the strategic impact of LFLs. This simplifying assumption can be relaxed in future work by testing alternative non-responsive profiles (e.g., seasonally adjusted, cluster-specific, etc).

\subsection{Continuous Action Space}

In the continuous-action setting, each LFL chooses a real-valued dispatch trajectory over the horizon subject to energy-balance and capacity constraints. At each update, an LFL computes its action by solving an optimization subproblem consistent with its response rule: either \emph{best-response dynamics} (BRD) or \emph{fictitious-play dynamics} (FPD). BRD updates myopically to the current condition, whereas FPD best-responds to a belief formed from the empirical average of others’ past actions, which tends to smooth reactions across time.

\subsubsection{Response Strategies: BRD vs.\ FPD} \quad

\begin{figure}[H]
\centering
\includegraphics[scale=0.55]{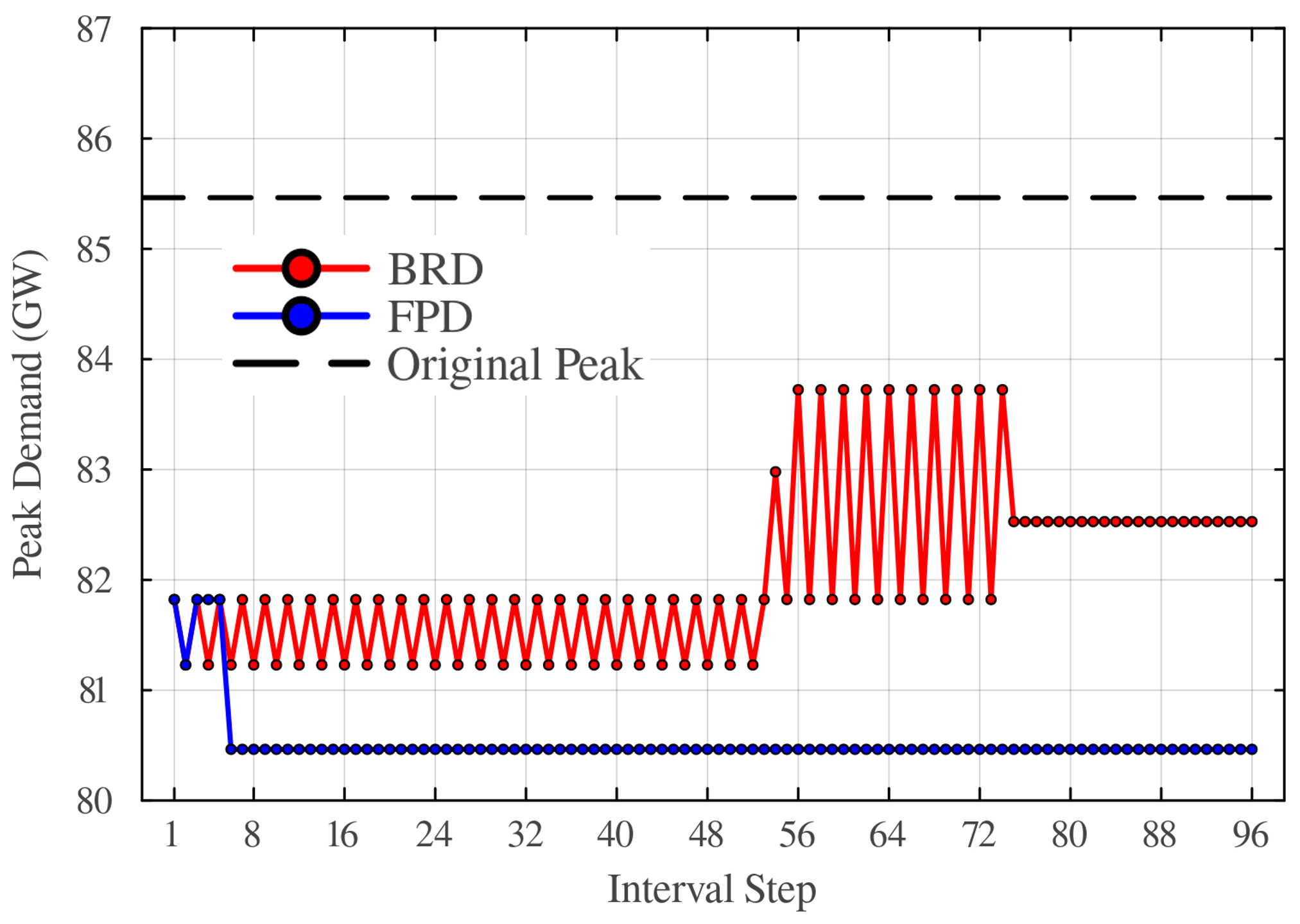}
\caption{Evolution of the system peak under BRD and FPD (continuous action space, \(N=5\)).}
\label{continuous_compare}
\end{figure}

\begin{figure}[H]
\centering
\includegraphics[scale=0.55]{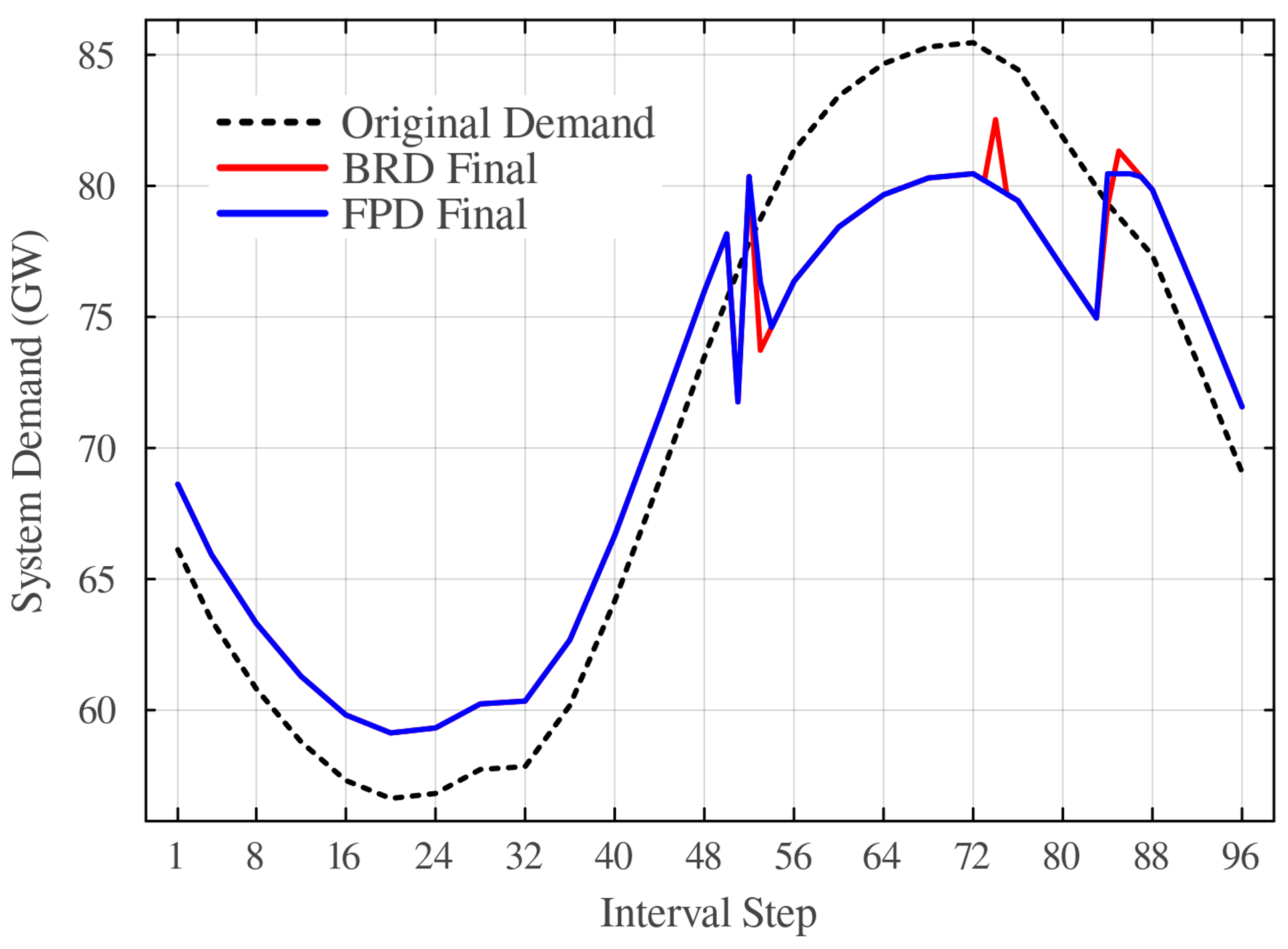}
\caption{Final demand curves under BRD and FPD (continuous action space, \(N=5\)).}
\label{continuous_final}
\end{figure}

To compare the dynamics induced by the two response rules, we consider \(N=5\) identical LFLs, each with an average demand of \(1000\)~MW and a capacity limit of \(\overline{X}_i = 1500\)~MW, corresponding to a 50\% operational margin relative to the average load. The real-time simulation horizon consists of 96 intervals, each representing a 15-minute time step. Figure~\ref{continuous_compare} illustrates the evolution of the system peak across these intervals under both BRD and FPD, while Figure~\ref{continuous_final} compares the resulting aggregate demand profiles at the end of the day.

\emph{Findings.} The result shows that FPD consistently lowers the peak and converges rapidly. In contrast, BRD is prone to oscillations and the creation of secondary peaks when multiple players react myopically during the peak hours.

\emph{Practical insight.} FPD differs from BRD in that each player best-responds to a belief formed from the \emph{empirical frequency} of others’ past actions, rather than to the instantaneous system profile. From an ISO’s perspective, broadcasting \emph{smoothed} CP-likelihood signals (rolling-average CP probabilities) can induce FPD-like behavior. Such coordination attenuates synchronized ``herding'' and improves peak outcomes relative to purely myopic, simultaneous best responses.

\subsubsection{Impact of Flexibility Level} \quad

We next fix \(N=5\) and examine BRD performance under three capacity caps representing different flexibility levels: \(\overline{X}_i \in \{1200,\,1500,\,1800\}\)~MW, corresponding to 20\%, 50\%, and 80\% margins relative to the \(1{,}000\)~MW average.

\begin{figure}[H]
\centering
\includegraphics[scale=0.56]{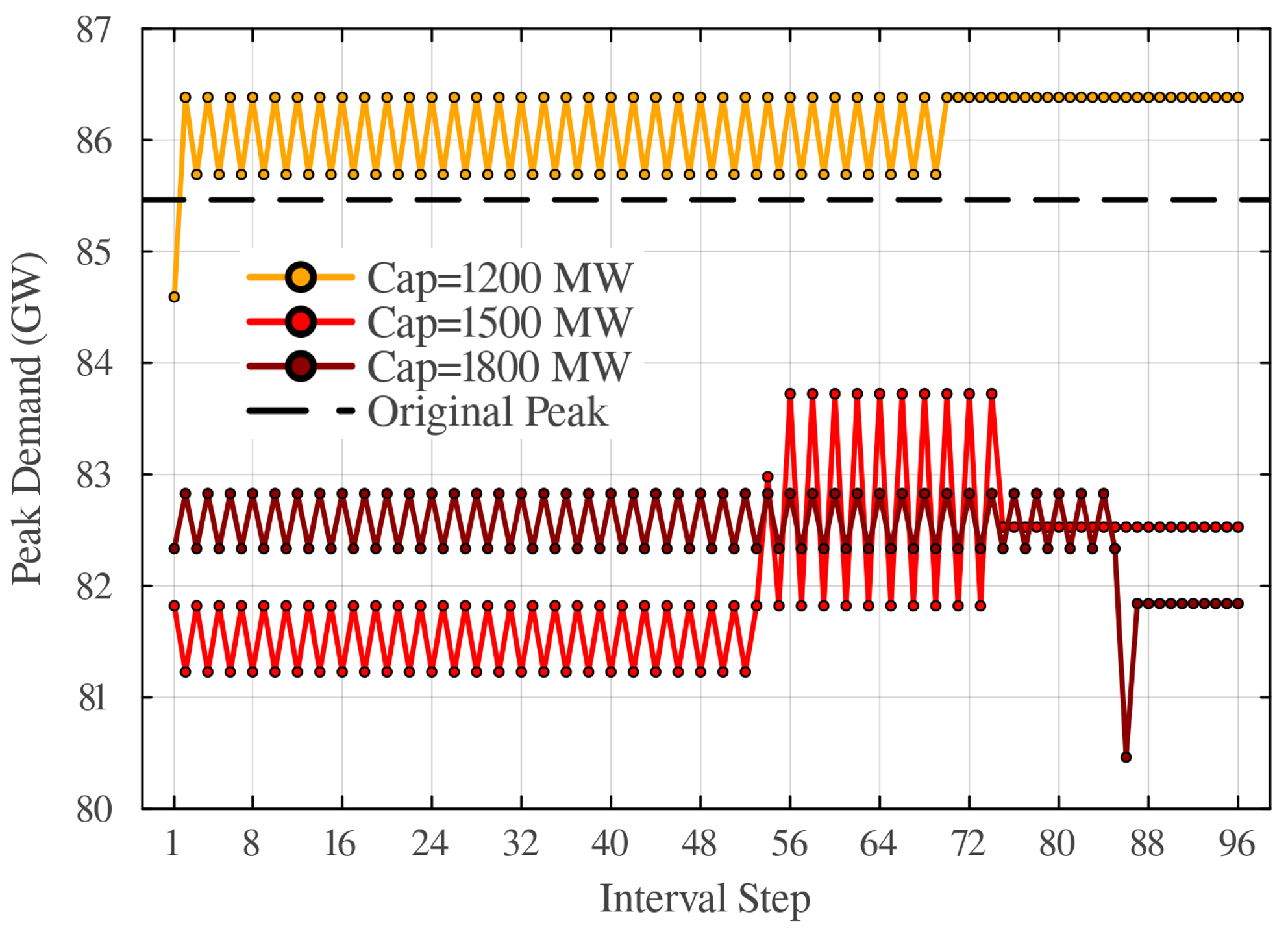}
\caption{BRD peak dynamics under different flexibility levels (\(\overline{X}_i=1200,\,1500,\,1800\)~MW, \(N=5\)).}
\label{continuous_caps}
\end{figure}

\emph{Findings.} Figure~\ref{continuous_caps} demonstrates that greater flexibility (larger \(\overline{X}_i\)) yields larger and more reliable peak reductions. By contrast, when capacity is tight (\(\overline{X}_i = 1200\)~MW, 20\% margin), BRD can \emph{backfire}: the \emph{final} system peak \emph{increases}, consistent with synchronized myopic shifts into the same candidate CP interval.

\emph{Practical insight.} When aggregate flexibility is scarce, myopic best responses tend to push load into the same candidate CP interval, amplifying the peak rather than shaving it. \emph{For participants}, reserving sufficient flexible capacity and distributing shifts across adjacent intervals reduces the likelihood of landing on the system peak. \emph{For system operators}, advising or requiring a minimum controllable-capacity threshold for LFL enrollment helps desynchronize responses and mitigates peak amplification.

\subsubsection{Impact of the Number of Participants} \quad

We vary the number of players from \(N=2\) to \(N=15\) while holding total responsive demand fixed (5000~MW) and comparing capacity caps $\overline{X}_i$ of 120\% and 180\% of the average.

\begin{figure}[H]
\centering
\includegraphics[scale=0.6]{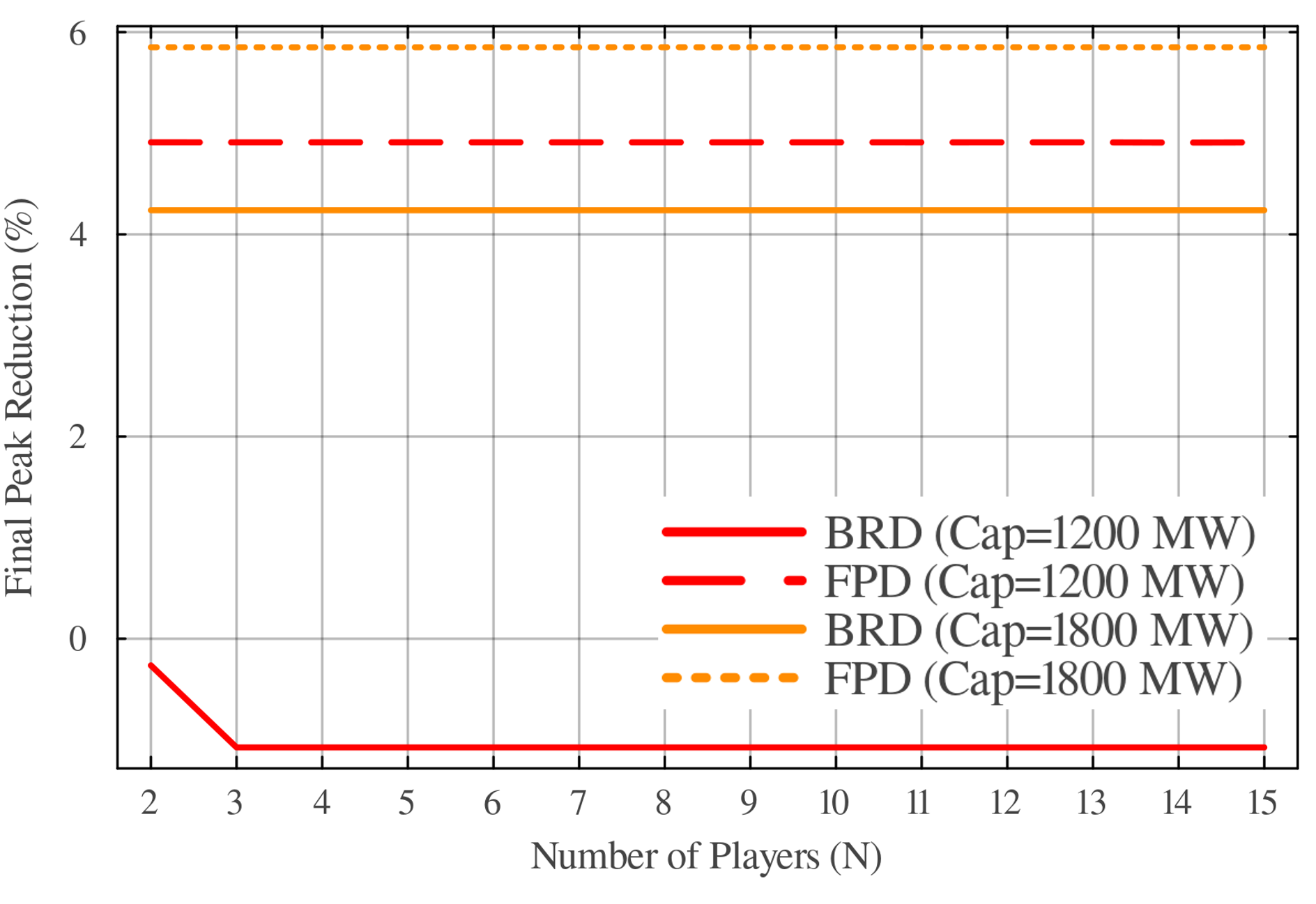}
\caption{Final peak reduction versus the number of players for different capacity caps (continuous action space).}
\label{continuous_players}
\end{figure}

\emph{Findings.} As shown in Figure~\ref{continuous_players}, when aggregate responsive capacity is held constant, the number of players has a limited effect on peak reduction. 

\emph{Practical insight.} At a fixed aggregate capacity, scaling from a few large to many small LFLs is largely neutral for peak shaving \emph{when agents share the same response rule and constraints}. 

\begin{table*}[b]
\centering
\caption{Peak Reduction (\%) Summary — Continuous Action Space}
\label{tab:cont-summary}

\renewcommand{\arraystretch}{1.2}   
\setlength{\tabcolsep}{8pt}         
\normalsize                         

\begin{tabular}{lcc}
\hline
\textbf{Metric} & \textbf{BRD} & \textbf{FPD} \\
\hline
Average over all Cases & 2.15 & 5.53 \\
Best Case & 5.42 ($\overline{X}_i{=}1500$, $N{=}14$) & 5.85 ($\overline{X}_i{=}1500 {~|~} 1800$, various $N$) \\
Worst Case & $-1.08$ ($\overline{X}_i$=1200, various $N$) & 4.91 ($\overline{X}_i$=1200, various $N$) \\
\hline
\end{tabular}
\end{table*}

\subsubsection{Summary Across Continuous-Action Scenarios} \quad

Table~\ref{tab:cont-summary} summarizes peak reductions across response rules (BRD, FPD), flexibility levels (\(1200/1500/1800\)~MW), and player counts (\(N=2\)–\(15\)).

On average, FPD achieves a \textbf{5.53\%} peak reduction, while BRD yields \textbf{2.15\%}. The best FPD cases reduce the peak by \textbf{5.85\%} ($\overline{X}_i$ \(=\) 1500 and 1800~MW across various \(N\)), whereas the worst BRD cases \emph{increase} the peak by \textbf{1.08\%} ($\overline{X}_i$ \(=\) 1200~MW across various \(N\)).

\emph{Key patterns and practical implications.}
\begin{enumerate}
\item \textbf{FPD dominates BRD across capacities and player counts.} Every FPD instance reduces the peak (4.9–5.9\%), highlighting the stabilizing role of belief averaging versus BRD’s myopic, potentially synchronized reactions.
\item \textbf{FPD’s peak shaving performance is robust to the number of players.} For each capacity level, FPD achieves nearly identical reductions regardless of \(N\) (e.g., \(\overline{X}_i=1{,}200\)~MW: \(4.91\%\); \(\overline{X}_i=1{,}500/1{,}800\)~MW: \(5.85\%\)). This indicates that FPD-style coordination scales well with participation and requires little tuning to population size.

\item \textbf{BRD is sensitive to flexibility and crowding.} With limited capacity ($\overline{X}_i$ \(=\) 1200~MW), BRD can backfire (up to +1.08\% peak). At higher capacity ($\overline{X}_i$ \(=\) 1800~MW), BRD becomes reliably beneficial (\(\approx\)4.24\%). Under mid-range capacity ($\overline{X}_i$ \(=\) 1500~MW), outcomes depend on \(N\) due to synchronization; coordination or randomized dispatch windows mitigate this risk.
\end{enumerate}

\subsection{Finite Action Space}

In the continuous formulation, each update under BRD or FPD requires solving a \emph{mixed-integer program} whose size grows with the length of the energy-balance window, which is technically solvable but can become time-consuming as the horizon expands~\cite{McCormick1976}. In practice, however, many LFLs operate with discrete setpoints or menu-based operating modes (due to equipment constraints, cycling limits, and operational policies). To reflect this reality and improve tractability, we consider a \emph{finite action space} in which each player selects from a predefined set of discrete load-shifting actions.

\emph{Experimental setup.}
We retain \(N=5\) identical players with capacity cap \(\overline{X}_i=1200\)~MW and down-sample the real-time horizon to 24 one-hour intervals (instead of 96 fifteen-minute intervals) to limit the combinatorial growth of the action set. We compare BRD and FPD under two discretizations: (i) a \emph{coarse} action space with relatively larger step sizes, and (ii) a \emph{fine-grained} action space with smaller step sizes. This design lets us isolate how action resolution interacts with learning dynamics to affect peak shaving.

The finite-action abstraction reduces each update to choosing from a menu rather than solving a full mixed integer programming problem, while capturing practical actuator limitations. 

\subsubsection{\textbf{Action Space 1 (Coarse)}}
We begin with a coarse discretization that captures simple ``on/off'' style operations. Each agent chooses from:
(i) a \emph{no-action} baseline, or
(ii) all combinations of \emph{complete shutdown} during any 1, 2, or 3 hours within the 24-hour horizon.
This yields 2324 feasible actions.
When an agent shuts down selected hours, the corresponding energy is reallocated to non–shutdown hours by filling the \emph{lowest-price} hours first, subject to per-interval capacity limits. Specifically, each non–shutdown hour can absorb at most \(200\)~MW of additional load because of \(\overline{X}_i=1200\)~MW.

\emph{Findings.} The dotted curves in Figure~\ref{Finite_space} (action space 1) show that BRD and FPD converge to \emph{similar final peaks}. The coarse menu restricts adjustment granularity, so both rules tend to concentrate shifts into the same low-price hours, limiting the scope for coordination gains.

\emph{Practical insight.}
Coarse, on/off–style control menus are easy to implement and compute, but can cap peak-shaving benefits because many agents select the same few “obvious” actions. 

\subsubsection{\textbf{Action Space 2 (Fine-Grained)}}
\begin{figure}[H]
\centering
\includegraphics[scale=0.46]{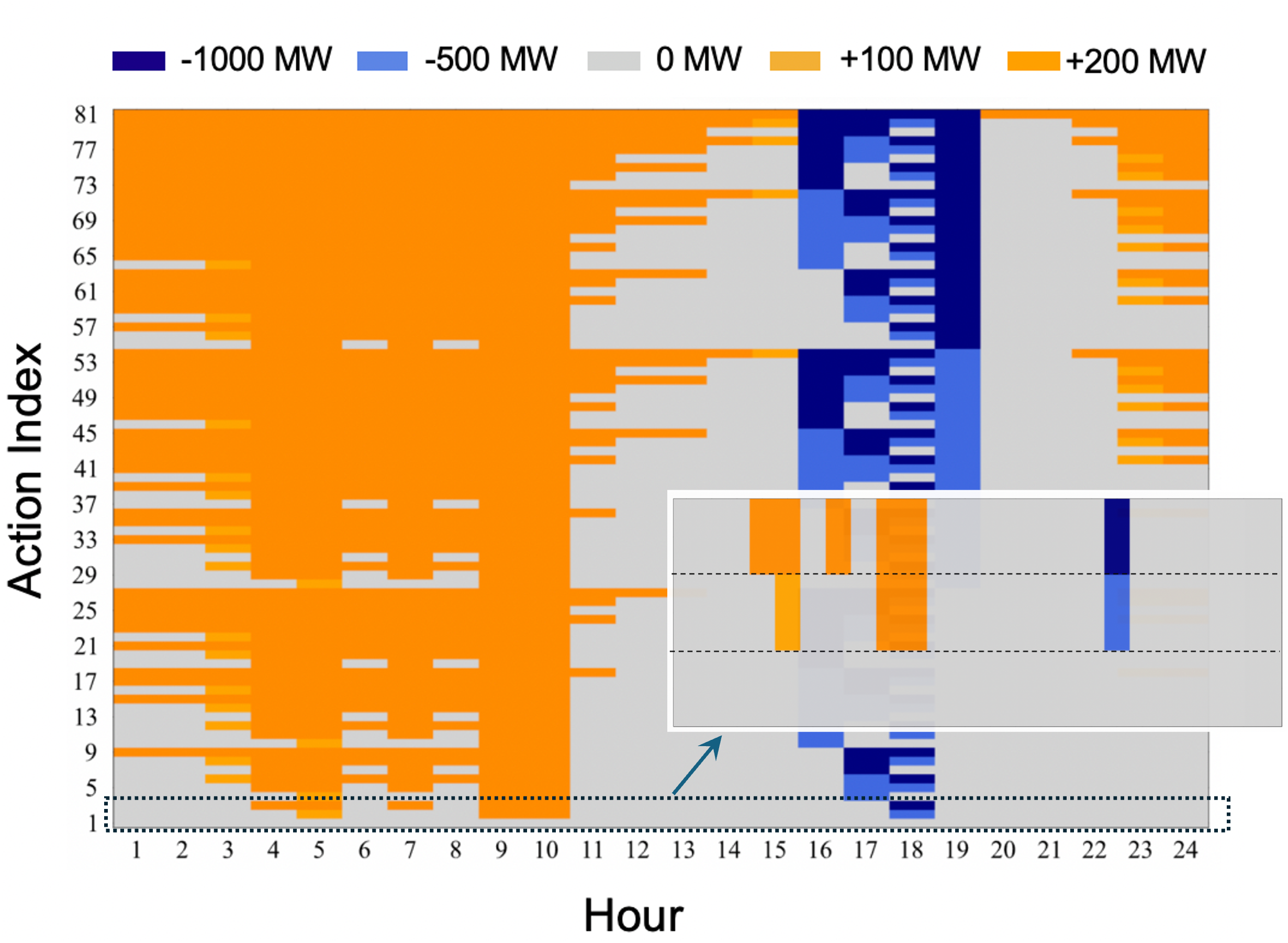}
\caption{Fine-grained action library for a single agent: each row is one action, with per-interval changes in \((-1000,-500,0,+200)\)~MW.}
\label{heatmap}
\end{figure}

To examine the value of resolution, we introduce a finer action space. Each agent first identifies the 4 highest-demand intervals in the baseline profile. It may then shift \emph{all} or \emph{half} of its demand away from any subset of these intervals to lower-price periods (again filled in ascending price order, respecting the \(+200\)~MW per-interval cap). Including the option to take no action, this construction yields \(|\mathcal{A}_i|=81\) discrete actions. Figure~\ref{heatmap} visualizes the action library: each row is a candidate action, and colors encode per-interval changes \((-1000,-500,0,+200)\)~MW. The inset also zooms into the first three actions.

\begin{figure}[H]
\centering
\includegraphics[scale=0.43]{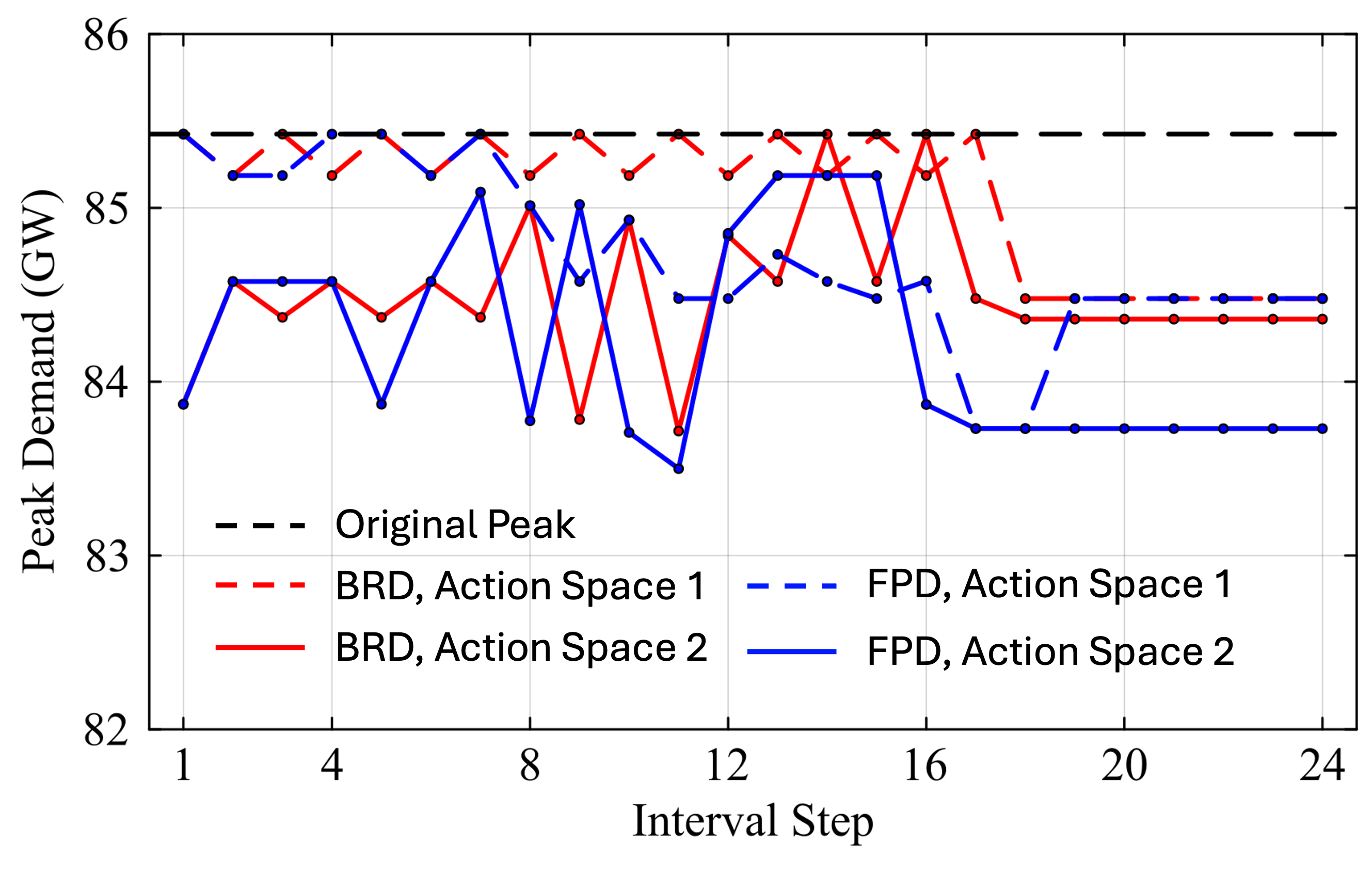}
\caption{Peak dynamics under coarse (dotted) and fine (solid) action spaces for BRD and FPD (\(N=5\), \(\overline{X}_i=1{,}200\)~MW, 24 intervals).}
\label{Finite_space}
\end{figure}

\emph{Findings.}
The solid curves in Figure~\ref{Finite_space} (action space 2) show that FPD achieves a \emph{larger} peak reduction than BRD under the same constraints. Moreover, for a fixed response rule, the fine-grained menu outperforms the coarse one: smaller step sizes diffuse shifts over more hours, reducing secondary-peak formation and oscillations.

\emph{Practical insight.}
Action \emph{resolution is flexibility}. \emph{For responsive consumers}, increasing the \emph{action-space resolution} improves peak-shaving performance and avoiding capacity charging. \emph{For ISOs}, incentivizing both \emph{flexible capacity} and \emph{control granularity}, coupled with smoothed CP-likelihood signals, encourages adoption, desynchronizes responses, and yields more reliable peak reductions.

\emph{Heterogeneity and scalability.} The benefit of finer action resolution can be interpreted as reducing discretization and avoiding the concentration of load shifts into a small set of intervals. The framework naturally accommodates heterogeneous participants through agent-specific capacity limits $\overline{X}_i$ and heterogeneous action sets $\mathcal{A}_i$. However, modeling strategic interactions among many heterogeneous agents can be computationally demanding. In large-scale settings, tractable approximations such as \emph{aggregation} (e.g., portfolio-level control for residential resources with device and comfort constraints), \emph{mean-field peak coupling}, and \emph{restricted (coarse-to-fine) action sets} preserve the CP peak interaction while reducing computational burden. This motivates the next subsection, where we study \emph{information providers} as a practical and scalable coordination mechanism.

\subsection{The Role of Information Provider}

The previous sections assume that participants independently observe others' actions and make decisions. In practice, system operators or third-party entities often provide forecasting signals or ``CP Alerts" to help users manage their exposure. We now evaluate the impact of  \emph{Information Providers} (IPs) that rank potential peak intervals and broadcast them to participants. 

Assuming IP identifies the top 8 candidate peak intervals for a particular day, we model agents’ responses using a simple yet practical operational rule \cite{gridstatus_ercot_4cp_2025_june}. Specifically, players curtail demand according to the IP’s ranking of these intervals:
\begin{itemize}
    \item \textbf{Top 2:} Full shutdown ($100\%$ reduction).
    \item \textbf{Rank 3:} Half shutdown ($50\%$ reduction).
    \item \textbf{Ranks 4--8:} Randomly select 3 intervals to shut down half ($50\%$).
\end{itemize}

Subject to the energy-balance constraint, the curtailed demand is reallocated to intervals with the lowest electricity prices, while respecting the capacity limit \(\overline{X}_i = 1200\)~MW. To account for the stochasticity induced by probabilistic shifting, the reported results are averaged over 20 independent and identically distributed simulation runs.

\subsubsection{IP's Ranking Strategies: Naive vs. Response-Aware} \quad

In practice, multiple IPs offer the ranking information in the market, often using different forecasting algorithms and ranking heuristics for candidate peak intervals. Here, we compare two representative ranking methodologies:
\begin{itemize}
\item \textbf{Naive IP:} Ranks the top 8 intervals based strictly on the original baseline demand profile.
\item \textbf{Response-Aware IP (Predictive):} Anticipates participant behavior. It calculates the system load profile \emph{assuming} all players respond to the Naive IP's advice, and then ranks the top 8 intervals of this projected ``post-response" profile.
\end{itemize}

\begin{figure}[H]
\centering
\includegraphics[scale=0.1]{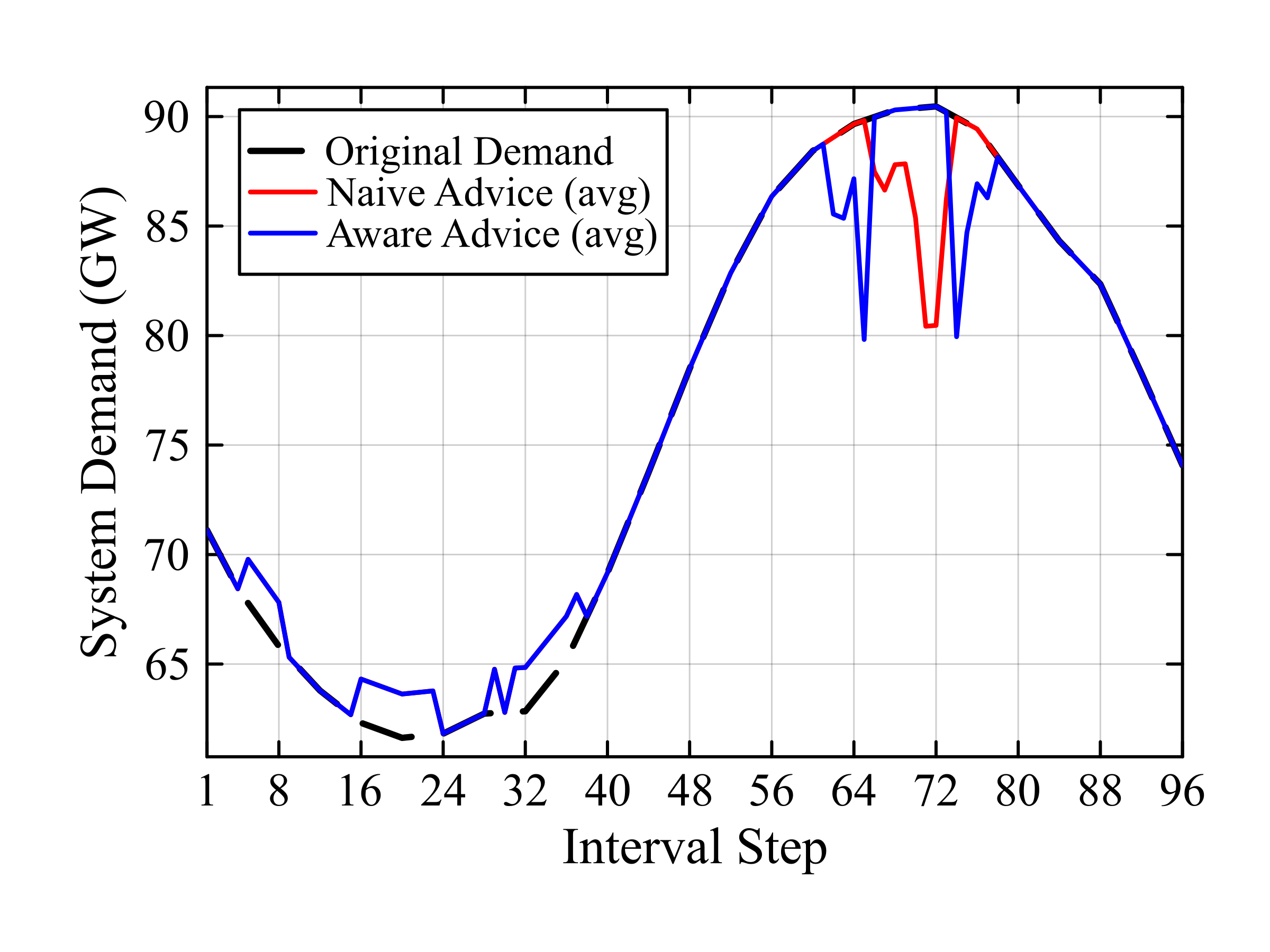}
\caption{System demand profiles under Naive (red solid) vs. Response-Aware (blue solid) Information Providers}
\label{ip_profiles}
\end{figure}

Figure~\ref{ip_profiles} compares the system load profiles before and after responses to information from different IPs. Although the Naive IP reduces the initial peak, the induced secondary peaks near its predicted peak intervals. By contrast, the Response-Aware IP anticipates the demand profile after the Naive-IP response and therefore targets intervals that \emph{would} become peaks, rather than the baseline peak.

The Naive and Response-Aware approaches represent two extreme cases of an information provider’s peak-ranking strategy. In practice, different participants may hold heterogeneous beliefs regarding the accuracy and credibility of the IP’s forecasts, leading to diverse response behaviors.

\subsubsection{Mixed Populations and Scalability} \quad

To assess the peak shaving value of different information providers, we assume that a subset of players follows the Naive IP’s ranking, while the remaining players adopt the Response-Aware IP’s ranking.

\begin{figure}[H]
\centering
\includegraphics[scale=0.1]{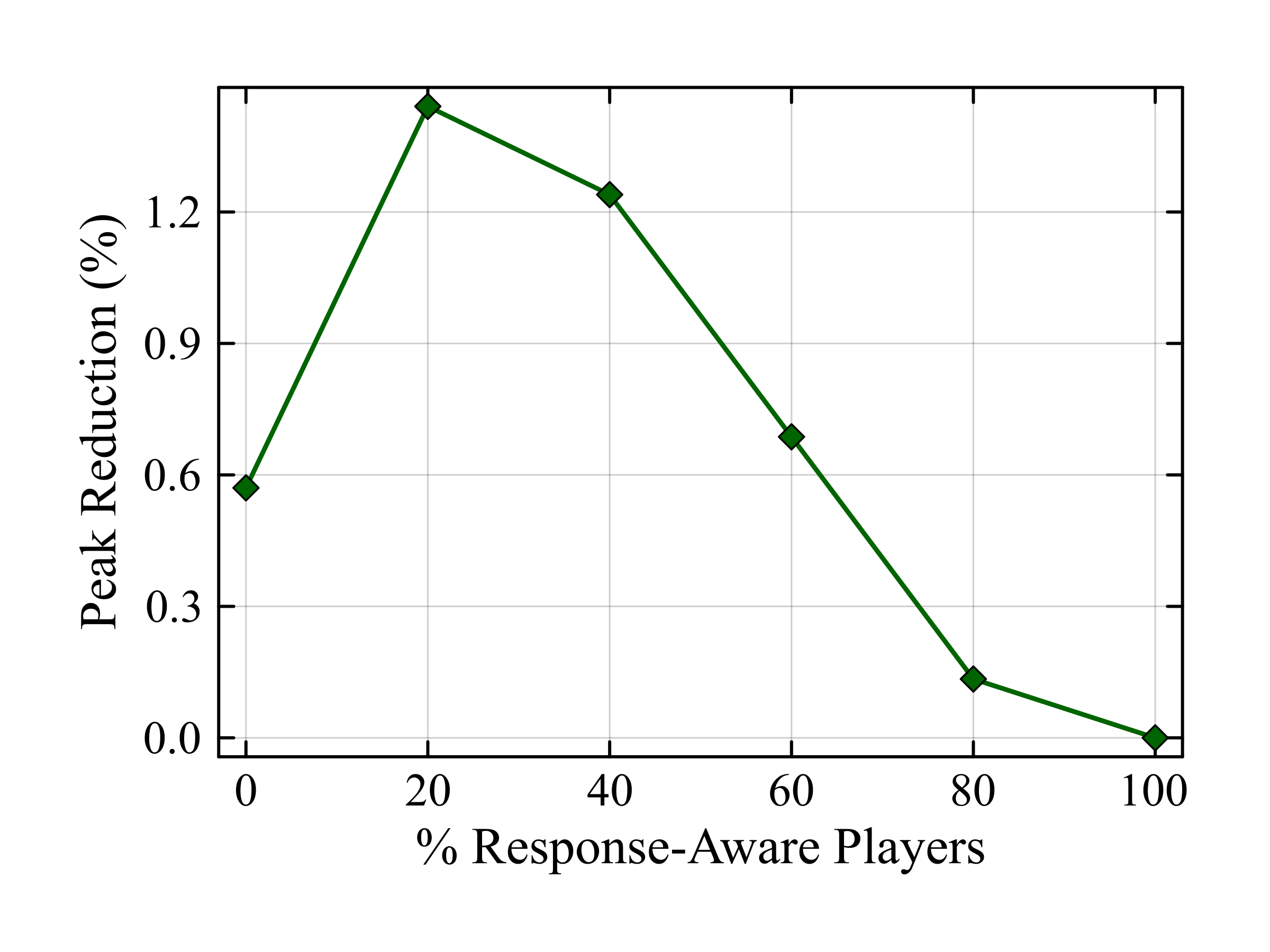}
\caption{Peak reduction sensitivity to population awareness level ($N=10$).}
\label{ip_mixed}
\end{figure}

\emph{The Value of a Mixed Population}. Figure~\ref{ip_mixed} reports the peak reduction as a function of the fraction of players following the Response-Aware advice. A key insight is the existence of an \emph{optimal population mix}. Intuitively, the Response-Aware cohort acts as a stabilizing buffer: by avoiding the herding points induced by the Naive ranking, these players redistribute load toward alternative intervals, thereby filling residual slack and mitigating the formation of secondary peaks. This result suggests that diversity in information providers and heterogeneous participant beliefs lead to more effective peak shaving than a setting in which all players follow a single common signal.

\begin{figure}[H]
\centering
\includegraphics[scale=0.1]{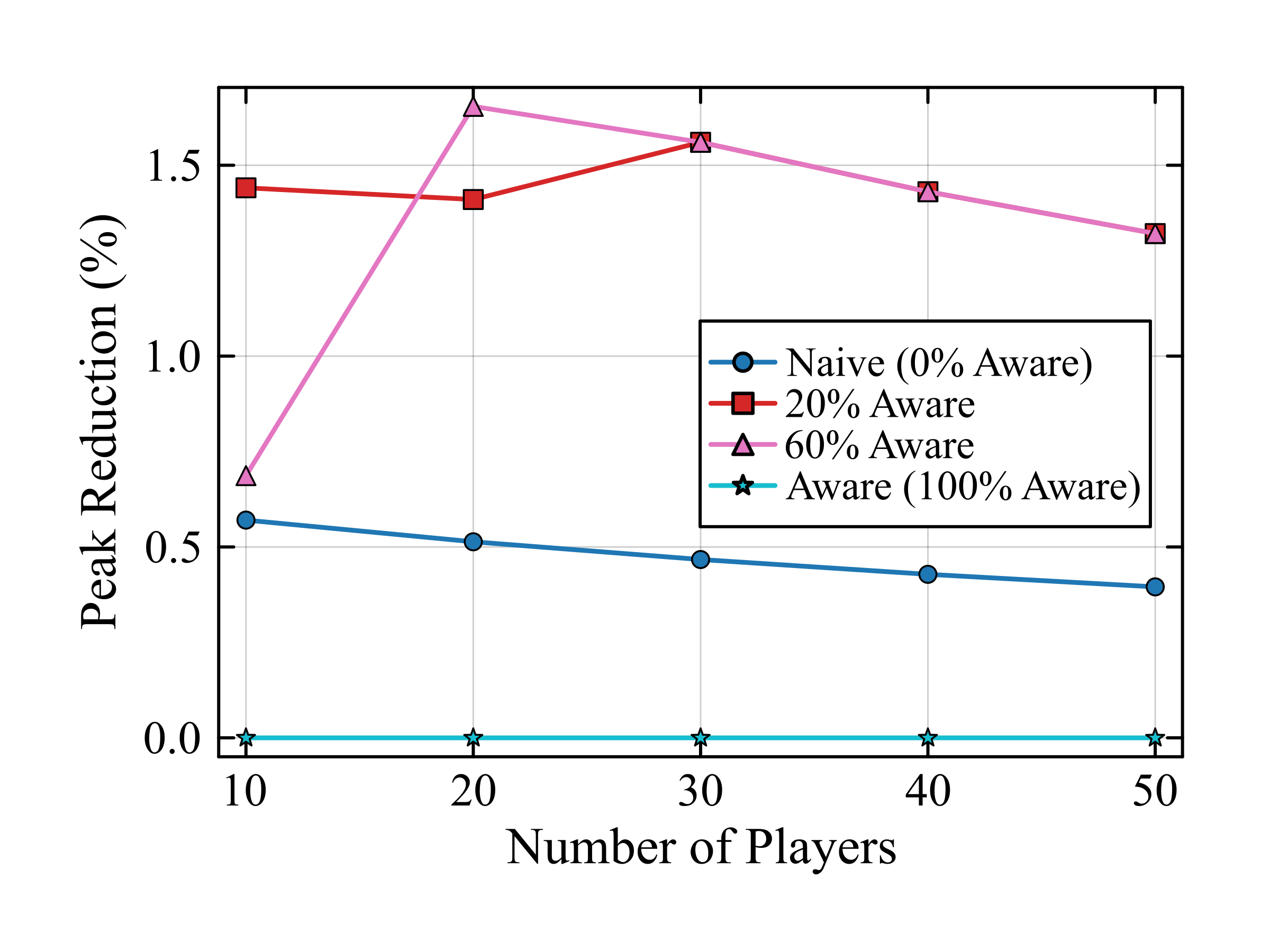}
\caption{The peak reduction under different numbers of players}
\label{ip_scaling}
\end{figure}

\emph{Scalability}. In the previous BRD and FPD cases, each player model (or estimate) the behavior of all other players, which becomes computationally challenging as the number of participants grows. The presence of information providers mitigates this burden by reducing the decision problem to responding to a ranked list of candidate peak intervals. As shown in Fig.~\ref{ip_scaling}, because all players apply the same response logic conditional on the IP signal, increasing the number of players has only a limited marginal effect on peak reduction. Nevertheless, scenarios with multiple information providers continue to outperform the single-provider case in terms of peak-shaving effectiveness.

\emph{Practical insight.} Coordination through information is a powerful and computationally lightweight alternative to fully specified dynamic-game models. Public load forecasts released by ISOs can be viewed as a Naive IP signal. By contrast, private IPs that incorporate response-aware peak forecasting can benefit their customers while also improving system-level peak shaving. As participation scales, however, the risk of \emph{asymmetric information} and \emph{information-induced herding} increases. Designing transparent and incentive-compatible mechanisms for such prediction services can therefore contribute to improved grid reliability \cite{chen2010designing}.

\section{Conclusion}

This paper examined the impact of CP pricing from behavioral and game-theoretic perspectives, linking the day-ahead one-shot game with rolling real-time decision making. We analyzed two canonical behavioral updates—best-response dynamics and fictitious-play dynamics—under both continuous and discrete action representations.

The results demonstrate that users’ belief formation and action flexibility are key determinants of peak-shaving effectiveness. Fictitious play yields stable and consistent peak reductions, whereas best-response dynamics is less reliable and can even increase peaks when flexibility is limited. Moreover, finer action resolution improves performance even when total controllable capacity is unchanged, while increasing the number of participants has limited marginal impact when aggregate flexibility is held constant.

We further show that information providers can influence system outcomes. While common signals may induce herding and secondary peaks, response-aware and heterogeneous signals mitigate this effect and improve peak-shaving performance. From a practical design standpoint, system operators can improve peak shaving effectiveness by enhancing probabilistic peak forecasting and setting minimum flexibility thresholds. For responsive consumers, maintaining adequate flexible capacity and adopting finer control options improve peak avoidance and contribute to system peak reduction. Future work will incorporate heterogeneous information and learning, network constraints, and the interaction between CP pricing, information providers, and overall market efficiency.

\section*{Disclaimer}
{The views expressed in this paper are the opinion of the
authors and do not reflect the views of PJM Interconnection,
L.L.C. or its Board of Managers, of which Le Xie is a member.}

\bibliographystyle{IEEEtran}
\bibliography{ref.bib}

\end{document}